\documentstyle[12pt,epsfig]{article}
\def\lbldef#1#2{\expandafter\gdef\csname #1\endcsname {#2}}

\def\href#1#2{#2}  

\begin{document}
\baselineskip=15.5pt
\pagestyle{plain}
\setcounter{page}{1}

\begin{titlepage}

\begin{flushright}
CERN-TH/2000-007\\
hep-th/0001018
\end{flushright}
\vspace{10 mm}

\begin{center}
{\Large Solitons in Brane Worlds II}

\vspace{5mm}

\end{center}

\vspace{5 mm}

\begin{center}
{\large Donam Youm\footnote{E-mail: Donam.Youm@cern.ch}}

\vspace{3mm}

Theory Division, CERN, CH-1211, Geneva 23, Switzerland

\end{center}

\vspace{1cm}

\begin{center}
{\large Abstract}
\end{center}

\noindent

We study the solution describing a non-extreme dilatonic $(p+1)$-brane 
intersecting a $D$-dimensional extreme dilatonic domain wall, where one 
of its longitudinal directions is along the direction transverse to the 
domain wall, in relation to the Randall-Sundrum type model.  The dynamics 
of the probe $(p+1)$-brane in such source background reproduces that of 
the probe $p$-brane in the background of the $(D-1)$-dimensional source 
$p$-brane.  However, as for a probe test particle, the dynamics in one 
lower dimensions is reproduced, only when the source $(p+1)$-brane is 
uncharged.  

\vspace{1cm}
\begin{flushleft}
CERN-TH/2000-007\\
January, 2000
\end{flushleft}
\end{titlepage}
\newpage

\section{Introduction}

Since it was shown \cite{rs} by Randall and Sundrum (RS) that 
gravity in the background of non-dilatonic domain wall with 
the exponentially decreasing warp factor is effectively compactified, 
some efforts have been made to understand gravitating objects living in 
such domain wall, e.g., Refs. \cite{gc,hwa,hm1,gs,hm2}.  We showed 
\cite{youm1} that dilatonic domain walls also effectively compactify gravity 
if the dilaton coupling parameter is sufficiently small.  The main motivation 
to consider dilatonic domain walls was that the consistency of equations 
of motion requires the cosmological constant term to have dilaton factor 
in order for the domain wall spacetime to admit {\it charged} brane 
solutions.  

In our previous work \cite{youm1}, we constructed completely localized 
solutions describing extreme charged branes living in the worldvolume 
of extreme dilatonic domain walls for the purpose of understanding charged 
branes in the RS type model.  Unexpectedly, it is found out \cite{youm2} 
that a charged $p$-brane is not effectively compactified to the 
charged $p$-brane in one lower dimensions.  It is speculated \cite{youm2} 
that this is due to the unusual properties of the Kaluza-Klein (KK) 
modes of gauge fields and also presumably form fields that the zero 
mode is not localized on the lower-dimensional hypersurface of the 
domain wall \cite{pom} and the massive modes strongly couple to the 
fields on the brane \cite{dhr,pom}.  So, following the result of Ref. 
\cite{hwa} that the Schwarzschild black hole in four-dimensional world 
within a domain wall should be regarded as an uncharged black string in 
five dimensions, we speculated \cite{youm2} that a charged $p$-brane in 
one lower dimensions might have to be regarded as a charged $(p+1)$-brane 
where one of its longitudinal directions is along the transverse direction 
of the domain wall.

It is the purpose of this paper to construct a solution describing a 
non-extreme charged dilatonic $(p+1)$-brane in an extreme dilatonic domain 
wall in $D$ dimensions where one of the longitudinal directions of the 
brane is along the transverse direction of the domain wall and to study 
its properties in relation to the RS type model.  We find that in the 
case of an uncharged branes, physics of the uncharged $p$-brane in one 
lower dimensions is reproduced by the uncharged $(p+1)$-brane in the 
domain wall.  However, when the $(p+1)$-brane is charged, the dynamics of 
a test particle in the background of the charged $p$-brane in one lower 
dimensions is not reproduced.  This is due to the fact that generally the 
transverse (to the domain walls) component of the spacetime metrics of 
charged branes in the domain walls has non-trivial dependence on the 
longitudinal coordinates of the domain walls.  Note, the original RS model 
\cite{rs2,rs,rs3} assumes that the perturbations of the domain wall 
metric should be along the longitudinal directions of the domain wall, 
only, in order for the lower-dimensional gravity to be reproduced.  So, 
in order for the RS type model to admit wide variety of gravitating 
objects which reproduce physics in one lower dimensions, one has to 
somehow modify the model.

The paper is organized as follows.  In section 2, we present the solution 
describing a nonextreme dilatonic $(p+1)$-brane intersecting an extreme 
dilatonic domain wall in $D$-dimensions where one of the longitudinal 
directions of the brane is along the direction transverse to the 
domain wall.  In section 2, we study the dynamics of the probe 
$(p+1)$-brane in such source background, comparing with the dynamics 
of the probe $p$-brane in the background of the $(D-1)$-dimensional 
source $p$-brane.  In section 3, we repeat the same analysis with a 
test particle.  The conclusion is given in section 4.

\section{Brane-World Solitons}

In this section, we discuss the $D$-dimensional solution describing a 
non-extreme dilatonic $(p+1)$-brane with the dilaton coupling 
parameter $a_{p+1}$ intersecting extreme dilatonic domain wall with 
the dilaton coupling parameter $a$ such that one of the longitudinal 
directions of the $(p+1)$-brane is along the direction transverse to 
the domain wall.  The configuration is given in the following table.  
\begin{center}
\begin{tabular}{|l||c|c|c|c|} \hline
{} \ & \ $t$ \ & \ ${\bf w}$ \ & \ ${\bf x}$ \ & \ $y$ 
\\ \hline\hline
brane \ & \ $\bullet$ \ & \ $\bullet$ \ & \ {} \ & \  $\bullet$ 
\\ \hline
domain wall \ & \ $\bullet$ \ & \ $\bullet$ \ & \ $\bullet$ \ & \ {}  
\\ \hline
\end{tabular}
\end{center}
Here, $t$ is the time coordinate, ${\bf w}=(w_1,...,w_p)$ and $y$ are 
the longitudinal coordinates of the $(p+1)$-brane, and ${\bf w}$ and 
${\bf x}=(x_1,...,x_{D-p-2})$ are the longitudinal coordinates of the 
domain wall.  The solution for such configuration solves the 
equations of motion of the following action:
\begin{equation}
S={1\over{2\kappa^2_D}}\int dx^D\sqrt{-g}\left[{\cal R}-{4\over{D-2}}
(\partial\phi)^2-{1\over{2\cdot(p+3)!}}e^{2a_{p+1}\phi}F^2_{p+3}
+e^{-2a\phi}\Lambda\right].
\label{einact}
\end{equation}
The solution has the following form:
\begin{eqnarray}
ds^2&=&H^{4\over{(D-2)\Delta}}\left[H^{-{{4(D-p-4)}\over{(D-2)
\Delta_{p+1}}}}_{p+1}\left(-fdt^2+dw^2_1+\cdots+dw^2_p\right)\right.
\cr
& &\left.+H^{{4(p+2)}\over{(D-2)\Delta_{p+1}}}_{p+1}\left(f^{-1}dx^2
+x^2d\Omega^2_{D-p-3}\right)\right]+H^{{4(D-1)}\over{(D-2)\Delta}}
H^{-{{4(D-p-4)}\over{(D-2)\Delta_{p+1}}}}_{p+1}dy^2,
\cr
e^{2\phi}&=&H^{{(D-2)a}\over{\Delta}}H^{{(D-2)a_{p+1}}\over
{\Delta_{p+1}}}_{p+1},
\cr
A_{tw_1...w_py}&=&{2\over\sqrt{\Delta_{p+1}}}{{\mu\cosh\delta_{p+1}
\sinh\delta_{p+1}}\over{x^{D-p-4}}}H^{-1}_{p+1}H^{4\over\Delta},
\label{pbrdwsol}
\end{eqnarray}
where the harmonic functions $H_{p+1}$ and $H$ for the $(p+1)$-brane 
and the domain wall, and the non-extremality function $f$ are given by
\begin{equation}
H_{p+1}=1+{{\mu\sinh^2\delta_{p+1}}\over{x^{D-p-4}}},\ \ \ \ \ \ \ 
H=1+Q|y|,\ \ \ \ \ \ \ 
f=1-{{\mu}\over{x^{D-p-4}}},
\label{harmfuncs}
\end{equation}
and the parameters $\Delta$'s in the solutions are defined as
\begin{eqnarray}
\Delta_{p+1}&=&{{(D-2)a^2_{p+1}}\over 2}+{{2(p+2)(D-p-4)}\over{D-2}},
\cr
\Delta&=&{{(D-2)a^2}\over{2}}-{{2(D-1)}\over{D-2}}.
\label{deltavals}
\end{eqnarray}
Here, $x\equiv|{\bf x}|$ is the radial coordinate of the transverse space 
of the $(p+1)$-brane, $\mu>0$ is the non-extremality parameter, and 
$Q$ is related to the cosmological constant $\Lambda$ as $\Lambda=
-2Q^2/\Delta$.  The extreme limit of the $(p+1)$-brane is achieved by 
taking $\mu\to 0$ such that $\mu e^{2\delta_{p+1}}$ is a non-zero constant.  
The consistency of the equations of motion requires that the dilaton 
coupling parameters satisfy the following constraint:
\begin{equation}
aa_{p+1}=-{{4(D-p-4)}\over{(D-2)^2}}.
\label{dilcplcnst}
\end{equation}
Note, this constraint (and the intersection rules arising from this 
constraint) is different from the ordinary one \cite{pt,tir,nf,ir} 
which is satisfied by the intersecting branes whose harmonic functions 
depend on the overall transverse coordinates.  The configuration under 
consideration in this paper does not have an overall transverse 
direction and each constituent is localized along the relative transverse 
directions.  So, the solution (\ref{pbrdwsol}) can be regarded as a 
particular case of general semi-localized solutions describing intersecting 
branes in which each constituent is localized along the relative transverse 
directions and delocalized along the overall transverse directions.  Such 
semi-localized intersecting brane solutions are constructed in Ref. 
\cite{ett} for the extreme case and in Ref. \cite{air} for the non-extreme 
case, and satisfy different intersection rules
\footnote{No-force requirement \cite{nf} on the probe brane in the source 
brane background yields the ordinary intersection rules, only.  So, the 
constraint (\ref{dilcplcnst}) on the dilaton coupling parameters is different 
from the one we expected through the no-force requirement in our previous 
work \cite{youm2}.}.  
An example is intersecting two NS5-branes with one-dimensional intersection 
\cite{khu,gkt}, rather than the three-dimensional one.  It is interesting 
to note that when $p=D-4$ there is no constraint on one of the dilaton 
coupling parameters if the other one is zero.  So, in this case, the bulk 
background of non-dilatonic domain walls ($a=0$) can admit charged 
$(p+1)$-branes with an arbitrary dilaton coupling parameter $a_{p+1}$.  
Also, in this case, the bulk background of both dilatonic ($a\neq 0$) and 
non-dilatonic ($a=0$) domain walls can admit non-dilatonic charged 
$(p+1)$-branes ($a_{p+1}=0$).  However, in this case the harmonic function 
$H_{p+1}$ is logarithmic.

\section{Probing Brane-World Solitons}

In this section, we repeat the probe dynamics analysis of our previous work 
\cite{youm2} with the source background (\ref{pbrdwsol}) of a brane-world 
soliton, comparing with the probe dynamics in the following source 
background of a $(D-1)$-dimensional non-extreme dilatonic $p$-brane: 
\begin{eqnarray}
ds^2&=&H^{-{{4(D-p-4)}\over{(D-3)\Delta_p}}}_p\left[-fdt^2+dw^2_1+\cdots+
dw^2_p\right]+H^{{4(p+1)}\over{(D-3)\Delta_p}}_p\left[f^{-1}dx^2+x^2
d\Omega^2_{D-p-3}\right],
\cr
e^{2\phi}&=&H^{{(D-3)a_p}\over{\Delta_p}}_p,\ \ \ \ \ \ \ \ \ 
A_{tx_1...x_p}={2\over{\sqrt{\Delta_p}}}{{\mu\cosh\delta_p\sinh
\delta_p}\over{x^{D-p-4}}}H^{-1}_p,
\label{pbrnsol}
\end{eqnarray}
where
\begin{eqnarray}
H_p&=&1+{{\mu\sinh^2\delta_p}\over{x^{D-p-4}}},\ \ \ \ \ \ \ 
f=1-{\mu\over{x^{D-p-4}}},
\cr
\Delta_p&=&{{(D-3)a^2_p}\over 2}+{{2(p+1)(D-p-4)}\over{D-3}}.
\label{harmdeldefs}
\end{eqnarray}
We regard this solution as being obtained by compactifying a 
$D$-dimensional $(p+1)$-brane with a dilaton coupling parameter $a_p$ 
(i.e., the solution (\ref{pbrdwsol}) with $H=1$) along a longitudinal 
direction (which is the $y$-direction in the notation of Eq. (\ref{pbrdwsol})) 
on $S^1$.  Since the parameter $\Delta_p$ is invariant under reductions or 
oxidations which do not involve field truncation, one can see that 
$a_p$ is related to $a_{p+1}$ of the solution in one higher dimension, 
namely that of the $(p+1)$-brane in Eq. (\ref{pbrdwsol}), as
\begin{equation}
(D-3)a^2_p=(D-2)a^2_{p+1}+4{{(D-p-4)^2}\over{(D-2)(D-3)}}.
\label{dilcppararel}
\end{equation}

\subsection{Probing with a $(p+1)$-brane}

The worldvolume action for a dilatonic $(p+1)$-brane with the following 
bulk action:
\begin{equation}
S_E={1\over{2\kappa^2_D}}\int d^Dx\sqrt{-g}\left[{\cal R}-{4\over{D-2}}
(\partial\phi)^2-{1\over{2\cdot (p+3)!}}e^{2a_{p+1}\phi}F^2_{p+3}\right]
\label{dilpbrnact}
\end{equation}
has the following form:
\begin{eqnarray}
S_{\sigma}&=&-T_{p+1}\int d^{p+2}\xi\left[e^{-a_{p+1}\phi}\sqrt{-{\rm det}\,
\partial_aX^{\mu}\partial_bX^{\nu}g_{\mu\nu}}\right.
\cr
& &\ \ \ \ \ \ \ \ \ \ \ \ \ \ \ 
\left.+{{\sqrt{\Delta_{p+1}}}\over{2}}{1\over{(p+2)!}}\epsilon^{a_1\dots 
a_{p+2}}\partial_{a_1}X^{\mu_1}
\dots\partial_{a_{p+2}}X^{\mu_{p+2}}A_{\mu_1\dots\mu_{p+2}}\right],
\label{wvdpbrnact}
\end{eqnarray}
where $T_{p+1}$ is the tension of the probe $(p+1)$-brane and the target 
space fields $g_{\mu\nu}$, $\phi$ and $A_{\mu_1\dots\mu_{p+2}}$ are the 
background fields (produced by the source brane) in which the probe 
$(p+1)$-brane with the target space coordinates $X^{\mu}(\xi^a)$ ($\mu=
0,1,...,D-1$) and the worldvolume coordinates $\xi^a$ ($a=0,1,...,p+1$) moves.

In the static gauge, in which $X^a=\xi^a$, the pull-back fields for the 
probe $(p+1)$-brane, oriented in the same way as the source $(p+1)$-brane, 
take the following forms:
\begin{eqnarray}
\hat{G}_{ab}&\equiv&g_{\mu\nu}\partial_aX^{\mu}\partial_bX^{\nu}=
g_{ab}+g_{ij}\partial_aX^i\partial_bX^j,
\cr
\hat{A}_{a_1...a_{p+2}}&\equiv&A_{\mu_1...\mu_{p+2}}\partial_{a_1}X^{\mu_1}
...\partial_{a_{p+2}}X^{\mu_{p+2}}=A_{a_1...a_{p+2}},
\label{pullbckst}
\end{eqnarray}
where the indices $i,j=1,...,D-p-2$ label the transverse space of the probe 
$(p+1)$-brane, i.e., $(X^i)=(x_1,...,x_{D-p-2})$ in the notation of Eq. 
(\ref{pbrdwsol}).  So, the worldvolume action (\ref{wvdpbrnact}) takes the 
following form:
\begin{equation}
S_{\sigma}=-T_{p+1}\int d^{p+2}\xi\left[e^{-a_{p+1}\phi}\sqrt{-{\rm det}
\left(g_{ab}+g_{ij}\partial_aX^i\partial_bX^j\right)}+{{\sqrt{\Delta_{p+1}}}
\over{2}}A_{01...p+1}\right].
\label{wvsimpact}
\end{equation}
From now on, we assume that the target space transverse coordinates 
$X^i$ for the probe $(p+1)$-brane depend on the time coordinate $\tau=\xi^0$ 
only, i.e., $X^i=X^i(\tau)$.

By substituting the explicit expressions (\ref{pbrdwsol}) for the source 
background fields of the brane-world $(p+1)$-brane into the general 
expression (\ref{wvsimpact}) for the probe $(p+1)$-brane action, one obtains 
the following:
\begin{eqnarray}
S_{\sigma}&=&-T_{p+1}\int d^{p+2}\xi\,H^{4\over\Delta}\left[H^{-1}_{p+1}
f^{1\over 2}\sqrt{1-H^{4\over{\Delta_{p+1}}}_{p+1}\left\{f^{-2}\left({{dx}
\over{d\tau}}\right)^2+x^2f^{-1}\mu^2_m\left({{d\phi_m}\over{d\tau}}\right)^2
\right\}}\right.
\cr
& &\ \ \ \ \ \ \ \ \ \ \ \ \ \ \ \ \ \ \ \ \ \ \ \ \ \ 
+\left.{{\mu\cosh\delta_{p+1}\sinh\delta_{p+1}}\over{x^{D-p-4}}}
H^{-1}_{p+1}\right].
\label{probp1bnact}
\end{eqnarray}
On the other hand, the probe $p$-brane action in the source 
background (\ref{pbrnsol}) of the $(D-1)$-dimensional $p$-brane is
\begin{eqnarray}
S_{\sigma}&=&-T_p\int d^{p+1}\xi\left[H^{-1}_pf^{1\over 2}\sqrt{
1-H^{4\over{\Delta_p}}_p\left\{f^{-2}\left({{dx}\over{d\tau}}\right)^2+
x^2f^{-1}\mu^2_m\left({{d\phi_m}\over{d\tau}}\right)^2\right\}}\right.
\cr
& &\ \ \ \ \ \ \ \ \ \ \ \ \ \ \ \ \ \ \ 
+\left.{{\mu\cosh\delta_p\sinh\delta_p}\over{x^{D-p-4}}}
H^{-1}_p\right].
\label{probpbnact}
\end{eqnarray}
Here, the angular coordinates $0\leq\phi_m<2\pi$ ($m=1,...,[(D-p-2)/2]$) 
are associated with $[(D-p-2)/2]$ rotation planes in the transverse space 
of the branes (with the coordinates ${\bf x}$) and the index $m$ is summed 
over $m=1,...,[(D-p-2)/2]$.  The remaining angular coordinates, which 
determine the direction cosines $\mu_m$, are constant due to the 
conservation of the direction of the angular momentum.

We see that the probe actions (\ref{probp1bnact}) and (\ref{probpbnact}) 
have the same form except that the probe $(p+1)$-brane action 
(\ref{probp1bnact}) has an additional overall factor $H^{4/\Delta}$.  
So, the dynamics of the probe $(p+1)$-brane in the background of the 
brane-world $(p+1)$-brane is identical to that of the probe $p$-brane 
in the background of the $p$-brane in one lower dimensions.  (Note, 
$\Delta_p=\Delta_{p+1}$, provided the $(D-1)$-dimensional $p$-brane is 
obtained by compactifying the $D$-dimensional $(p+1)$-brane on $S^1$ 
without field truncation.)  The effect of the overall factor $H^{4/\Delta}$ 
in the former case is to effectively increase [decrease] the tension of the 
probe $(p+1)$-brane when $\Delta>0$ [$\Delta<0$], namely $T^{\rm eff}_{p+1}
=T_{p+1}H^{4/\Delta}$.  

Since the probe actions for the two cases have the same form, one also 
expects that the first law of black brane thermodynamics of the $p$-brane 
in one lower dimensions can be extracted from that of the $(p+1)$-brane 
in the domain wall, and vice versa.  The first law of black brane 
thermodynamics of the latter brane with the solution given by Eq. 
(\ref{pbrdwsol}) is
\begin{equation}
\delta M_{p+1}=T^{p+1}_H\delta S_{p+1}+\Phi_{p+1}\delta Q_{p+1},
\label{1stlaw}
\end{equation}
where $M_{p+1}$ and $S_{p+1}$ are the ADM mass and the entropy of the source 
$(p+1)$-brane per unit $(p+1)$-brane worldvolume, $Q_{p+1}$ is the source 
$(p+1)$-brane charge normalized to take integer values (i.e., the number of 
elementary $(p+1)$-branes with unit charge), and the Hawking temperature 
$T^{p+1}_H$ and the chemical potential $\Phi_{p+1}$ of the source 
$(p+1)$-brane are given by
\begin{equation}
T^{p+1}_H={{D-p-4}\over{4\pi\mu^{1\over{D-p-4}}\cosh^{4\over{\Delta_{p+1}}}
\delta_{p+1}}},\ \ \ \ \ \ 
\Phi_{p+1}={2\over\sqrt{\Delta_{p+1}}}T_{p+1}H^{4\over\Delta}
\tanh\delta_{p+1}.
\label{temcp}
\end{equation}
In the case of the source $p$-brane in $(D-1)$-dimensions with the solution 
given by Eq. (\ref{pbrnsol}), the temperature $T^p_H$ and the chemical 
potential $\Phi_p$ are respectively related to those of the $(p+1)$-brane 
as $T^p_H=T^{p+1}_H$ and $\Phi_p/T_p=\Phi_{p+1}/(T_{p+1}H^{4/\Delta})$, if 
we let $\delta_p=\delta_{p+1}$.  Note, $\Delta_{p+1}$ in Eq. (\ref{pbrdwsol}) 
and $\Delta_p$ in Eq. (\ref{pbrnsol}) are the same, if the $p$-brane 
is obtained from the $(p+1)$-brane through the dimensional reduction 
without field truncation.  One can think of the changes $\delta M_{p+1}$, 
$\delta S_{p+1}$ and $\delta Q_{p+1}$ as being due to an addition of the probe 
$(p+1)$-brane to the source $(p+1)$-brane \cite{kt}.  Namely, we bring the 
probe $(p+1)$-brane with a unit brane charge from spatial infinity 
($x=\infty$) to the source brane horizon ($x=x_H=\mu^{1/(D-p-4)}$).  
Then, one can interpret $T^{p+1}_H\delta S_{p+1}$ as the heat released by 
the probe $(p+1)$-brane while it falls inside the source $(p+1)$-brane, 
which is just the difference in static potential energy $V_{p+1}(x)$ of 
the probe $(p+1)$-brane, i.e., $T^{p+1}_H\delta S_{p+1}=V(\infty)-V(x_H)$ 
\cite{kt}.   From the above probe actions (\ref{probp1bnact}) and 
(\ref{probpbnact}), one obtains the following static potentials on the 
probe branes:
\begin{equation}
V_{p+1}=T_{p+1}H^{4\over\Delta}\left(f^{1\over 2}+{{\mu\cosh\delta_{p+1}
\sinh\delta_{p+1}}\over{x^{D-p-4}}}\right)H^{-1}_{p+1},
\label{statpot1}
\end{equation}
for the probe $(p+1)$-brane, and
\begin{equation}
V_p=T_p\left(f^{1\over 2}+{{\mu\cosh\delta_p\sinh\delta_p}\over{x^{D-p-4}}}
\right)H^{-1}_p,
\label{statpot2}
\end{equation}
for the probe $p$-brane.  As expected, in the extreme limit ($\mu\to 0$ 
with $\mu e^{2\delta}$ finite constant), the potentials are constant in 
accordance with the no-force condition for extreme branes.  We see 
that the two static potentials are related as $V_{p+1}/(T_{p+1}
H^{4/\Delta})=V_p/T_p$ and therefore $T^{p+1}_H\delta S_{p+1}/(T_{p+1}
H^{4/\Delta})=T^p_H\delta S_{p}/T_p$, if $\delta_p=\delta_{p+1}$.  
The probe actions (\ref{probp1bnact}) and (\ref{probpbnact}) with 
$\delta_p=\delta_{p+1}$ imply that the mass density changes are 
related as $\delta M_{p+1}/(T_{p+1}H^{4/\Delta})=\delta M_p/T_p$.  
Since the probe branes have unit charges, $\delta Q_p=1=\delta 
Q_{p+1}$.  Gathering all the above, one can bring the first law of the 
black $(p+1)$-brane thermodynamics (\ref{1stlaw}) to the following 
first law of thermodynamics of the black $p$-brane in $D-1$ dimensions:
\begin{equation}
\delta M_p=T^p_H\delta S_p+\Phi_p\delta Q_p.
\label{1stlawpbrn}
\end{equation}

\subsection{Probing with a test particle}

In analyzing the dynamics of a test particle in a curved spacetime 
background, it is convenient to utilize the symmetry of the spacetime.  
The Killing vectors of the spacetime metrics of both of the solutions 
(\ref{pbrdwsol}) and (\ref{pbrnsol}) are $\partial/\partial t$, 
$\partial/\partial w_i$ and $\partial/\partial\phi_m$.  Contracting these 
Killing vectors with the velocity $U^{\mu}=dx^{\mu}/d\lambda$ of the test 
particle along the geodesic path $x^{\mu}(\lambda)$ parametrized by an 
affine parameter $\lambda$, one obtains the following constants of motion 
for the test particle:
\begin{eqnarray}
E&=&-g_{\mu\nu}\left({{\partial}\over{\partial t}}\right)^{\mu}U^{\nu}
=-g_{tt}{{dt}\over{d\lambda}},
\cr
p^i&=&g_{\mu\nu}\left({{\partial}\over{\partial w_i}}\right)^{\mu}U^{\nu}
=g_{ii}{{dw_i}\over{d\lambda}},
\cr
J^m&=&g_{\mu\nu}\left({{\partial}\over{\partial \phi_m}}\right)^{\mu}U^{\nu}
=g_{\phi_m\phi_m}{{d\phi_m}\over{d\lambda}}.
\label{cstmtn}
\end{eqnarray}
In addition, there is another constant of motion associated with metric 
compatibility along the geodesic path:
\begin{equation}
\epsilon=-g_{\mu\nu}{{dx^{\mu}}\over{d\lambda}}{{dx^{\nu}}\over
{d\lambda}},
\label{metcomp}
\end{equation}
where $\epsilon=1,0$ respectively for a massive particle (i.e., a 
timelike geodesic) and a massless particle (i.e., a null geodesic).  

For the simplicity of the calculation, we shift the transverse coordinate 
$y$ of the domain wall so that the harmonic function for the domain wall 
takes the form $H=Qy$, where we restrict to the region $y\geq 0$.  Then, 
we apply the following change of coordinate:
\begin{equation}
y=\left({{\Delta+2}\over\Delta}Q^{-{2\over\Delta}}z
\right)^{\Delta\over{\Delta+2}}
\label{chcoordyz}
\end{equation}
to bring the domain wall metric to the conformally flat form.  In 
this new coordinate, the metric in Eq. (\ref{pbrdwsol}) takes the 
following form:
\begin{eqnarray}
ds^2&=&\left({{\Delta+2}\over\Delta}Qz\right)^{{4}\over{(D-2)(\Delta+2)}}
\left[H^{-{{4(D-p-4)}\over{(D-2)\Delta_{p+1}}}}_{p+1}\left(-fdt^2+dw^2_1
+\cdots+dw^2_p\right)\right.
\cr
& &\left.+H^{{4(p+2)}\over{(D-2)\Delta_{p+1}}}_{p+1}\left(f^{-1}dx^2
+x^2d\Omega^2_{D-p-3}\right)+H^{-{{4(D-p-4)}\over{(D-2)\Delta_{p+1}
}}}_{p+1}dz^2\right].
\label{confflatmet}
\end{eqnarray}

For the test particle moving along the $z$-direction, i.e., only the 
$z$-component of $U^{\mu}$ is non-zero, the geodesic motion is described 
by the following equation, derived from the geodesic equation and Eq. 
(\ref{metcomp}):
\begin{equation}
{d\over{d\lambda}}\left[z^{4\over{(D-2)(\Delta+2)}}{{dz}\over{d\lambda}}
\right]=-{{2\epsilon}\over{(D-2)(\Delta+2)}}\left({{\Delta+2}\over\Delta}
Q\right)^{-{4\over{(D-2)(\Delta+2)}}}H^{{4(D-p-4)}\over{(D-2)
\Delta_{p+1}}}_{p+1}{1\over z}.
\label{zdireceq}
\end{equation}
The geodesic path $z(\lambda)$ for a massless test particle 
(i.e., the $\epsilon=0$ case) is $z={\rm constant}$ or
\begin{equation}
z=z_0\lambda^{{(D-2)(\Delta+2)}\over{(D-2)(\Delta+2)+4}},
\label{nullgeod}
\end{equation}
where $z_0$ is an arbitrary constant.   The general explicit expression for 
the timelike geodesic path (i.e., the $\epsilon=1$ case) is hard to obtain.  
So, in the following we shall consider only the case of the null geodesic 
motion along the $z$-direction.  The null geodesic path $z={\rm constant}$ 
simply corresponds to the motion constrained along the longitudinal 
directions of the domain wall.  

Making use of the constants of the motion in Eq. (\ref{cstmtn}), one can 
put Eq. (\ref{metcomp}) into the following form:
\begin{equation}
\left({{dx}\over{d\lambda}}\right)^2+{{g_{zz}}\over{g_{xx}}}\left({{dz}
\over{d\lambda}}\right)^2+{{J^2_m}\over{g_{xx}g_{\phi_m\phi_m}}}+
{{E^2}\over{g_{xx}g_{tt}}}+{\epsilon\over{g_{xx}}}=0,
\label{geoeqn}
\end{equation}
where the index $m$ is summed over $m=1,...,[(D-p-2)/2]$.  The angular 
coordinates of the transverse space of the branes (with the coordinates 
${\bf x}$) in the spherical coordinates, except for the ones $\phi_m$ 
associated with the angular momenta $J_m$, are constant due to the 
conservation of the direction of the angular momentum.  And we are considering 
the motion of the test particle with the longitudinal coordinates $w_i$ 
of the branes constant, which is possible due to the conservation of the 
linear momenta $p^i$ along those directions.  By plugging the explicit 
expression (\ref{confflatmet}) for the background metric of the source 
brane into the general expression (\ref{geoeqn}), we obtain
\begin{eqnarray}
\left({{dx}\over{d\lambda}}\right)^2&+&{\cal C}^{-2}\left[\left\{
{\cal C}^2H^{-{4\over{\Delta_{p+1}}}}_{p+1}\left({{dz}\over{d\lambda}}
\right)^2+{{\cal J}^2\over{H^{{8(p+2)}\over{(D-2)\Delta_{p+1}}}_{p+1}x^2}}
\right\}f-E^2H^{{4(D-2p-6)}\over{(D-2)\Delta_{p+1}}}_{p+1}\right]
\cr
&+&\epsilon_{p+1}{\cal C}^{-1}H^{-{{4(p+2)}\over{(D-2)
\Delta_{p+1}}}}_{p+1}f=0,
\label{geomotn1}
\end{eqnarray}
where ${\cal C}=\left({{\Delta+2}\over\Delta}Qz\right)^{4\over
{(D-2)(\Delta+2)}}$ is the conformal factor in the metric (\ref{confflatmet}), 
$z=z(\lambda)$ for the null geodesic motion is constant or is given by 
Eq. (\ref{nullgeod}), and $\epsilon_{p+1}=0,1$ respectively for the 
massless and the massive test particle.  On the other hand, for the 
test particle in the source background (\ref{pbrnsol}) of the 
$(D-1)$-dimensional $p$-brane, the geodesic motion is described by
\begin{equation}
\left({{dx}\over{d\lambda}}\right)^2+\left[\epsilon_pH^{-{{4(p+1)}\over
{(D-3)\Delta_p}}}_p+{{\cal J}^2\over{H^{{8(p+1)}\over{(D-3)\Delta_p}}_px^2}}
\right]f-E^2H^{{4(D-2p-5)}\over{(D-3)\Delta_p}}_p=0,
\label{geomotn2}
\end{equation}
where $\epsilon_p=1,0$ respectively for the timelike and the null 
geodesic.
Here, ${\cal J}$ in the above is defined in terms of the 
conserved angular momenta $J^m$ of the test particle as
\begin{equation}
{\cal J}^2\equiv \sum^{[{{D-p-2}\over 2}]}_{m=1}{{(J^m)^2}\over{\mu^2_m}},
\label{jdef}
\end{equation}
where the direction cosines $\mu_m$ specifying the direction of $x$ are 
constant due to conservation of the direction of angular momentum (therefore 
${\cal J}$ is also constant).  

First, when the $(p+1)$-brane is uncharged (i.e., $H_{p+1}=1$), one can 
bring the equation (\ref{geomotn1}) for the null geodesic motion 
($\epsilon_{p+1}=0$) to the form of the equation (\ref{geomotn2}) for 
the time-like geodesic motion ($\epsilon_p=1$) in the $(D-1)$-dimensional 
uncharged $p$-brane background.  To see this, we consider the following 
equation obtained from Eq. (\ref{geomotn1}) by setting $\epsilon_{p+1}=0$ 
and $H_{p+1}=1$, and substituting (\ref{nullgeod}):
\begin{eqnarray}
\left({{dx}\over{d\lambda}}\right)^2+\left({{\Delta+2}\over\Delta}Qz_0
\right)^{-{8\over{(D-2)(\Delta+2)}}}\lambda^{-{8\over{(D-2)(\Delta+2)+4}}}
\left[\left\{\left({{\Delta+2}\over\Delta}Qz_0\right)^{8\over{(D-2)(\Delta+2)}}
\right.\right.
\cr
\left.\left.\times\left({{(D-2)(\Delta+2)z_0}\over{(D-2)(\Delta+2)+4}}
\right)^2+{{\cal J}^2\over{x^2}}\right\}\left(1-{\mu\over{x^{D-p-4}}}
\right)-E^2\right]=0.
\label{nullgeodex}
\end{eqnarray}
By redefining the radial coordinate, constants of the motion and parameters 
in the following way:
\begin{eqnarray}
\tilde{x}&\equiv&{\cal A}x,\ \ \ \ \ \ \ \ 
\tilde{E}\equiv {\cal A}E,\ \ \ \ \ \ \ \ 
\tilde{J}\equiv {\cal A}^2J,\ \ \ \ \ \ \ \ 
\tilde{\mu}\equiv {\cal A}^{D-p-4}\mu,
\cr
\tilde{\lambda}&\equiv&{{(D-2)(\Delta+2)+4}\over{(D-2)(\Delta+2)}}
\left({{\Delta+2}\over\Delta}Qz_0\right)^{-{2\over{(D-2)(\Delta+2)}}}
\lambda^{{(D-2)(\Delta+2)}\over{(D-2)(\Delta+2)+4}},
\label{redef}
\end{eqnarray}
where
\begin{equation}
{\cal A}\equiv \left({{(D-2)(\Delta+2)z_0}\over{(D-2)(\Delta+2)+4}}
\right)^{-1}\left({{\Delta+2}\over\Delta}Qz_0\right)^{-{4\over{(D-2)
(\Delta+2)}}},
\label{defred}
\end{equation}
one can bring Eq. (\ref{nullgeodex}) into the following form:
\begin{equation}
\left({{d\tilde{x}}\over{d\tilde{\lambda}}}\right)^2+
\left(1+{\tilde{\cal J}^2\over\tilde{x}^2}\right)
\left(1-{\tilde{\mu}\over\tilde{x}^{D-p-4}}\right)=\tilde{E}^2.
\label{nulltotime}
\end{equation}
This reproduces the equation for the timelike geodesic motion in the 
background of uncharged $(D-1)$-dimensional $p$-brane, i.e., Eq. 
(\ref{geomotn2}) with $\epsilon_p=1$ and $H_p=1$.  This result generalizes 
the result of Ref. \cite{hwa} to the case of an uncharged black 
brane in a dilatonic domain wall in arbitrary spacetime dimensions.  

Next, we consider the charged branes.  The equation for the null geodesic 
motion (with nontrivial lightlike motion along the $z$-direction) in the 
background of the brane-world charged $(p+1)$-brane, i.e., Eq. 
(\ref{geomotn1}) with $\epsilon_{p+1}=0$ and Eq. (\ref{nullgeod}) substituted, 
reduces to the following form after the quantities are redefined as in Eq. 
(\ref{redef}):
\begin{equation}
\left({{d\tilde{x}}\over{d\tilde{\lambda}}}\right)^2+
\left[\tilde{H}^{-{4\over{\Delta_{p+1}}}}_{p+1}-{\tilde{\cal J}^2\over
{\tilde{H}^{{8(p+2)}\over{(D-2)\Delta_{p+1}}}_{p+1}\tilde{x}^2}}\right]
\tilde{f}-\tilde{E}^2\tilde{H}^{{4(D-2p-6)}\over{(D-2)\Delta_{p+1}}}_{p+1}=0,
\label{chgeonull}
\end{equation}
where
\begin{equation}
\tilde{H}_{p+1}=1+{{\mu\sinh^2\delta_{p+1}}\over{({\cal A}
\tilde{x})^{D-p-4}}},\ \ \ \ \ 
\tilde{f}=1-{\tilde{\mu}\over\tilde{x}^{D-p-4}}.
\label{modhramdefs}
\end{equation}
This equation is different from the equation for the timelike 
geodesic in the background of the $(D-1)$-dimensional $p$-brane, 
i.e., Eq. (\ref{geomotn2}) with $\epsilon_p=1$, since the powers of the 
harmonic functions are different in the two equations.  This difference 
might be attributed to the fact that when one compactifies the 
Einstein-frame metric for the $D$-dimensional $(p+1)$-brane (i.e., Eq. 
(\ref{confflatmet}) without the $z$-dependent conformal factor) along 
one of its longitudinal directions (i.e., the $z$-direction) by using the 
KK metric Ansatz without the Weyl-scaling factor in the $(D-1)$-dimensional 
part of the metric (i.e., $g_{\mu\nu}={\rm diag}(\bar{g}_{\bar{\mu}\bar{\nu}},
\bar{\varphi})$ with $\mu,\nu=0,1,...,D-1$ and $\bar{\mu},\bar{\nu}=0,1,...,
D-2$), one gets non-Einstein-frame metric for the $(D-1)$-dimensional 
$p$-brane.  However, as can be seen from Eq. (\ref{geoeqn}), even in such 
non-Einstein-frame spacetime in $D-1$ dimensions, the equation for the 
timelike geodesic will look different because of the dependence of the 
$(z,z)$-component of the $D$-dimensional metric on $x$.  (Cf. The second 
term on the LHS of Eq. (\ref{geoeqn}) for the null geodesic motion in $D$ 
dimensions is identified with the last term on the LHS of Eq. (\ref{geoeqn}) 
for the timelike geodesic motion in $D-1$ dimensions.)  We just write down 
the equation for the timelike geodesic motion in such $(D-1)$-dimensional 
background for comparison:
\begin{equation}
\left({{dx}\over{d\lambda}}\right)^2+\left[H^{-{{4(p+2)}\over
{(D-2)\Delta_{p+1}}}}_{p+1}+{{\cal J}^2\over{H^{{8(p+2)}\over{(D-2)
\Delta_{p+1}}}_{p+1}x^2}}\right]f-E^2H^{{4(D-2p-6)}\over{(D-2)
\Delta_{p+1}}}_{p+1}=0.
\label{timegeod2}
\end{equation}
As mentioned, the first terms in the square brackets of Eqs. 
(\ref{chgeonull}) and (\ref{timegeod2}) are different.  So, only when the 
probe motion along the $z$-direction is trivial, i.e., $z={\rm constant}$, 
the null geodesic motion in the background of the $D$-dimensional 
$(p+1)$-brane (described by Eq. (\ref{geomotn1}) with $\epsilon_{p+1}=0$ 
and $z={\rm constant}$) reproduces the null geodesic motion in the 
background of the $p$-brane in one lower dimensions (described by Eq. 
(\ref{geomotn2}) with $\epsilon_p=0$).  In the case of the uncharged branes, 
the above problems did not arise because the $(z,z)$-component of the 
$D$-dimensional metric is independent of $x$ and the dimensional reduction 
of the Einstein-frame metric for the $D$-dimensional uncharged $(p+1)$-brane 
along a longitudinal direction (without the Weyl-scaling term in the metric) 
leads to the Einstein-frame metric for the $(D-1)$-dimensional uncharged 
$p$-brane.   It is interesting to note that for the $p=D-4$ case the equation 
(\ref{chgeonull}) for the null geodesic motion (with nontrivial motion along 
the $z$-direction) in the background of the brane-world $(p+1)$-brane 
reproduces the equation for the timelike geodesic motion in the background 
of the $p$-brane in one lower dimensions, i.e., Eq. (\ref{geomotn2}) with 
$\epsilon_p=1$ and Eq. (\ref{timegeod2}).  In such case, the transverse (to 
the domain wall) component of the metric (\ref{confflatmet}), i.e., the 
$(z,z)$-component, is independent of the longitudinal coordinates of the 
domain wall.   This is in accordance with our speculation on the source of 
disparity in the probe particle dynamics in the backgrounds of {\it charged} 
branes.

In fact, one of the assumptions of the RS model is that the $D$-dimensional 
conformally flat metric (of the domain wall) should have the perturbation 
around the flat metric along the longitudinal directions of the domain 
wall, only.  Indeed, as mentioned in the above, one can see from Eq. 
(\ref{geoeqn}) that the extra space component of the metric, i.e., $g_{zz}$, 
should be independent of the longitudinal coordinates of the domain wall, 
in order for the null geodesic motion (with non-trivial lightlike motion 
along the $z$-direction) in $D$ dimensions to reproduce the timelike geodesic 
motion in the background of the source $p$-brane in $D-1$ dimensions.  
So, for the $p=D-4$ case, even if the branes are charged, the null 
geodesic motion in the source background of the brane-world $(p+1)$-brane 
reproduces the timelike geodesic motion in the source background of the 
$p$-brane in one lower dimensions, because the extra space component $g_{zz}$ 
of the brane-world $(p+1)$-brane metric is independent of the longitudinal 
coordinates of the domain wall.  Also, recently, it is shown \cite{iv} that 
the metric perturbation in the direction transverse to the domain wall is 
not localized on the brane.  On the other hand, in general, the extra space 
component $g_{zz}$ of the spacetime metrics for charged brane solutions in 
the domain wall depends on the longitudinal coordinates ${\bf x}$ of 
the domain wall.  So, it seems inevitable that all the charged branes in 
brane worlds do not reproduce physics in one lower dimensions.  This might 
be the indication of either the need to modify physics in one lower 
dimensions (which seems unlikely) or the limitation of the current RS model 
that needs to be modified so that it can accommodate, for example, charged 
black holes or branes that will reproduce lower-dimensional physics.  
(Also, the non-dilatonic domain wall of the RS model in Refs. 
\cite{rs2,rs,rs3} does not admit even charged black string solutions, 
which are supposed to be identified as charged black holes in 
four dimensions, because of the constraint on the dilaton coupling 
parameters.)  

\section{Conclusion}

We studied the non-extreme dilatonic $(p+1)$-brane in the bulk of extreme 
dilatonic domain wall, where one of the longitudinal directions of the 
brane is along the transverse direction of the wall.  Such $(p+1)$-brane 
is expected to be the domain wall bulk counterpart to the ordinary $p$-brane 
observed on the hypersurface of the domain wall.  We studied the probe 
dynamics on such background for the purpose of seeing whether the effective 
compactification of such $(p+1)$-brane through the RS type domain wall 
leads to the physics of $p$-brane in one lower dimensions obtained through 
the ordinary KK compactification or not.  In this paper, we found partial 
agreement of the probe dynamics in the two backgrounds.  Namely, in the case 
of the probe $(p+1)$-brane in the background of the source $(p+1)$-brane in 
the bulk of the domain wall, its dynamics reproduces the dynamics of the 
probe $p$-brane in the background of the source $p$-brane originated from the 
source $(p+1)$-brane (without the domain wall) through the ordinary KK 
compactification.  However, in the case of the test particle moving in the 
same source backgrounds, we found agreement only for the case when the source 
branes are uncharged.  We have attributed this difference to the fact that 
the metric for the charged brane solutions does not satisfy the RS gauge 
condition.   Namely, RS showed \cite{rs} that the gravity in one lower 
dimensions is reproduced only for the case where the domain wall metric has 
perturbations along the longitudinal directions of the wall, only, whereas 
the transverse (to the wall) component of the metric for charged branes in 
the bulk of domain walls in general has non-trivial dependence on the domain 
wall worldvolume coordinates.  Perhaps, in some cases RS type models might 
give rise to the effective theory in one lower dimensions different from the 
one that would have been obtained through the ordinary KK compactification.  
On the other hand, the RS gauge for the domain wall metric perturbation used 
in Ref. \cite{rs} is applicable only for the case when there are no 
additional fields in the action.  Namely, when additional fields are added 
to the action, one might have to modify the RS gauge condition of Ref. 
\cite{rs}, possibly, to include the transverse perturbation as well.  
Furthermore, the trick used in studying the geodesic motion of the test 
particle with non-trivial motion along the extra space, which was devised 
in Ref. \cite{hwa} and also applied in this paper, rather seems not to be 
rigorous.  Definite answer to this point cannot be given until one would 
solve the full coupled nonlinear geodesic equations $\textnormal{\it\" 
x}^{\mu}+\Gamma^{\mu}_{\nu\rho}\dot{x}^{\nu}\dot{x}^{\rho}=0$.  We defer 
the above unanswered questions to our future work.

\end{document}